\documentclass[11pt,a4paper]{article}
\usepackage[utf8]{inputenc}
\usepackage{amsmath}
\usepackage{amsfonts}
\usepackage{authblk}
\usepackage{amssymb}
\usepackage{subcaption}
\usepackage{graphicx}
\usepackage[left=1.00cm, right=1.00cm, top=1.00cm, bottom=1.00cm]{geometry}
\usepackage[colorlinks = true, linkcolor = blue, urlcolor  = blue, citecolor = blue, anchorcolor = blue]{hyperref}

\author{T. Kiran and M. Ponmurugan \footnote{email correspondence: ponphy@cutn.ac.in}}
\affil{Department of Physics, School of Basic and Applied Sciences, Central University of Tamilnadu, Thiruvarur - 610 005, Tamilnadu, India}
\date{}
\title{ The invariant-based shortcut to adiabaticity for qubit heat engine operates under quantum Otto cycle}
\begin{document}
	\maketitle
\subparagraph{Abstract:}In this paper, we study the role and relevance of the cost for an invariant-based shortcut to adiabaticity enabled qubit heat engine operates in a quantum Otto cycle. We consider a qubit heat engine with Landau-Zener Hamiltonian and improve its performance using the Lewis-Riesenfeld invariant-based shortcut to adiabaticity method. Addressing the importance of cost for better performance, the paper explores its relationship with the work and efficiency of the heat engine. We analyze the cost variation with the time duration of non-adiabatic unitary processes involved in the heat engine cycle. The cost required to attain the quasi-static performance of the qubit heat engine is also discussed. We found the efficiency lost due to non-adiabaticity of the engine can be revived using the shortcut method and it is even possible to attain quasi-static performance under finite time with higher cost.  
\section{Introduction}
\label{Introduction}
 A large fraction of quantum heat engine studies explored various quantum mechanical analogs of classical thermodynamic cycles~\cite{QuanQuantum2007,QuanQuantum2009,ZagoskinSqueezing2012,YuQuantum2017,KosloffThe2017}, the engines with different quantum mechanical systems as the working substance~\cite{AlickiThe1979,LindenHow2010,JianhuiQuantum2012,BenentiFundamental2017}, the time duration of thermodynamic processes~\cite{GevaA1992,ZhengOccurrence2016,SelcukIrreversible2017,PezzuttoAn2019,SangyunFinite2020} and the performance of the quantum heat engines~\cite{AbahSingle2012,DiazQuantum2014,LeeNonuniversality2018,DeffnerEfficiency2018,DorfmanEfficiency2018,VarinderPerformance2020}. Among the quantum analogs of classical thermodynamic cycles, Quantum Otto Cycle (QOC) is the most explored and inspected with various working mediums. QOC consists of two quantum isochoric and two quantum adiabatic processes. Ideally, the quantum adiabatic processes are slow processes under isolation resulting in an engine with vanishing power output. The slowness of the adiabatic process is required to preserve the initial quantum states' occupation probability in the final state of the working medium. On the other hand, finite time driving will cause quantum non-adiabatic transitions even under isolation. Thus the initial probability of occupation of quantum states is altered during the finite time non-adiabatic processes. The Shortcut to Adiabaticity (STA) methods~\cite{BerryTransitionless2009,TorronteguiShortcuts2013,CampoShortcuts2013,DeffnerClassical2014} can suppress these quantum non-adiabatic transitions to execute the QOC in the finite time duration.

STA processes are generally non-adiabatic processes that reproduce the adiabatic final state in a finite time duration~\cite{CampoShortcuts2019}. It is both theoretically and experimentally shown that the STA can improve the quantum heat engines' performance~\cite{DengBoosting2018,AbahPerformance2018,AbahShortcut2019}. There are several ways to achieve STA including, counter-diabatic driving~\cite{BerryTransitionless2009,TakahashiTransitionless2013,FunoUniversal2017,BarisFInite2021}, local counter-diabatic driving~\cite{IbezMultiple2012}, Lewis-Riesenfeld (LR) invariant based inverse engineering~\cite{LewisAn1969,ChenFast2010,TorronteguiFast2011,ChenLewis2011,ChenOptimal2011,TorronteguiFast2012}, and a few more~\cite{OdelinShortcuts2019}. The LR invariant based STA method involves the reconstruction of the externally controllable parameter of the Hamiltonian from an arbitrary time-dependent parameter called the scaling factor~\cite{LewisAn1969}. This inverse engineering should be following the preassigned boundary conditions to assure the final adiabatic state. Instead of inverse engineering the Hamiltonian, we use the Hermitian invariant itself to execute STA with appropriate boundary conditions. In other words, the Hermitian LR invariant takes over the non-adiabatic expansion and compression processes of QOC from the Hamiltonian. The advantage of this method is that we can derive an LR invariant with a convenient number of control parameters. Thus, not only the scaling factor is preassigned, but all other arbitrary parameters of the invariant can be designed for our purpose. The recently introduced parameter ‘cost’ for the realization of Shortcut methods accounts for the use of resources in its operation~\cite{CampoShortcuts2019,ZhengCost2016,CampbellTrade2017}. Recent studies are concentrating on the cost of an STA protocol for a range of time durations \cite{BarisFInite2021,AbahEnergetic2019,CakmakSpin2019}. The output of such studies manifests the cost as a function of the time duration of the process. However, In this study, we use the freedom to arbitrarily vary the invariant's parameters to change the cost of implementation and discuss the cost required for a particular STA protocol in a predetermined time duration.

The working substance is another crucial element in designing a quantum heat engine. The qubit is a two-level quantum system that mainly finds application in quantum information and quantum computation~\cite{NielsenQuantum2000}. The heat engines with two-level quantum systems as the working substance (hereafter we call it Qubit Heat Engine (QHE)) is a topic of current theoretical and experimental focus in quantum thermodynamics~\cite{RobsnagelA2019,PetersonExperimental2019}. The methods to suppress the non-adiabaticity in QHE are also taken into consideration using STA with counter-diabatic driving~\cite{CakmakSpin2019}. In general, the qubit Hamiltonian of different instants of time does not commute with each other. Then, the fundamental quantum mechanics necessitate the time ordering of evolution during non-adiabatic expansion and compression processes. In general, the invariant of such a Hamiltonian also shows the non-commutation property. Hence, the time ordering of the evolution during the STA process has to take into account. The time ordering of evolution during an invariant-based STA process is yet to consider in literature. In our previous paper~\cite{KiranInvariant2021}, we have considered the quantum oscillator under a time-dependent frictional contact and applied the STA with invariant-based inverse engineering to regain the ideal harmonic oscillator dynamics. In general, a time-dependent harmonic oscillator Hamiltonian at an instant of time may not commute with the Hamiltonian of other instances. However, we did not consider the time-ordered evolution of the oscillator since STA based on the conventional inverse engineering approach disregards the importance of time-ordered evolution. In a recent paper~\cite{SolfanelliNonadiabatic2020}, the authors considered the time-ordered evolution for a non-adiabatic quantum Otto cycle and investigated its possibilities as a heat engine, refrigerator, heater, and thermal accelerator, obeying the Clausius inequality for cyclic processes. They characterized the QOC using the heat exchanged during the cycle's isochoric branches under the assumption of weak qubit heat bath coupling. Using the theoretical model provided in Ref.~\cite{SolfanelliNonadiabatic2020}, we study the STA-enabled non-adiabatic QOC with time-ordered evolution during expansion and compression processes. 

In this paper, we provide the general conditions for the possible operation of QOC as a non-adiabatic heat engine with qubit as its working substance. We follow the same method in Ref.~\cite{SolfanelliNonadiabatic2020} to characterize the QHE using the heat exchanged in the isochoric processes under the assumption of weak qubit heat bath coupling. We provide all the general formulations in section~\ref{NonAQOC}. Section~\ref{QHELZ} incorporates the formulation developed in section \ref{NonAQOC} for a non-adiabatic QHE with Landau-Zener (LZ) Hamiltonian and analyzes its performance in terms of work and efficiency. The above analysis identify the time durations with a diminished performance of non-adiabatic QHE. In section~\ref{STAI}, we discuss the STA mechanism using the LR invariant in the above framework to improve the QHE performance. The performance improvement is examined concerning the cost for implementation of the protocol. The observations demand a minimum cost to achieve better performance than the non-adiabatic QHE. Also, the cost is finitely higher to obtain the quasi-static characteristics within a predetermined time duration. We conclude the paper with section~\ref{Conclusion}.
 
\section{Non-adiabatic quantum Otto cycle}
\label{NonAQOC}
\begin{figure}[h]
	\centering
	\includegraphics[width=.6\linewidth]{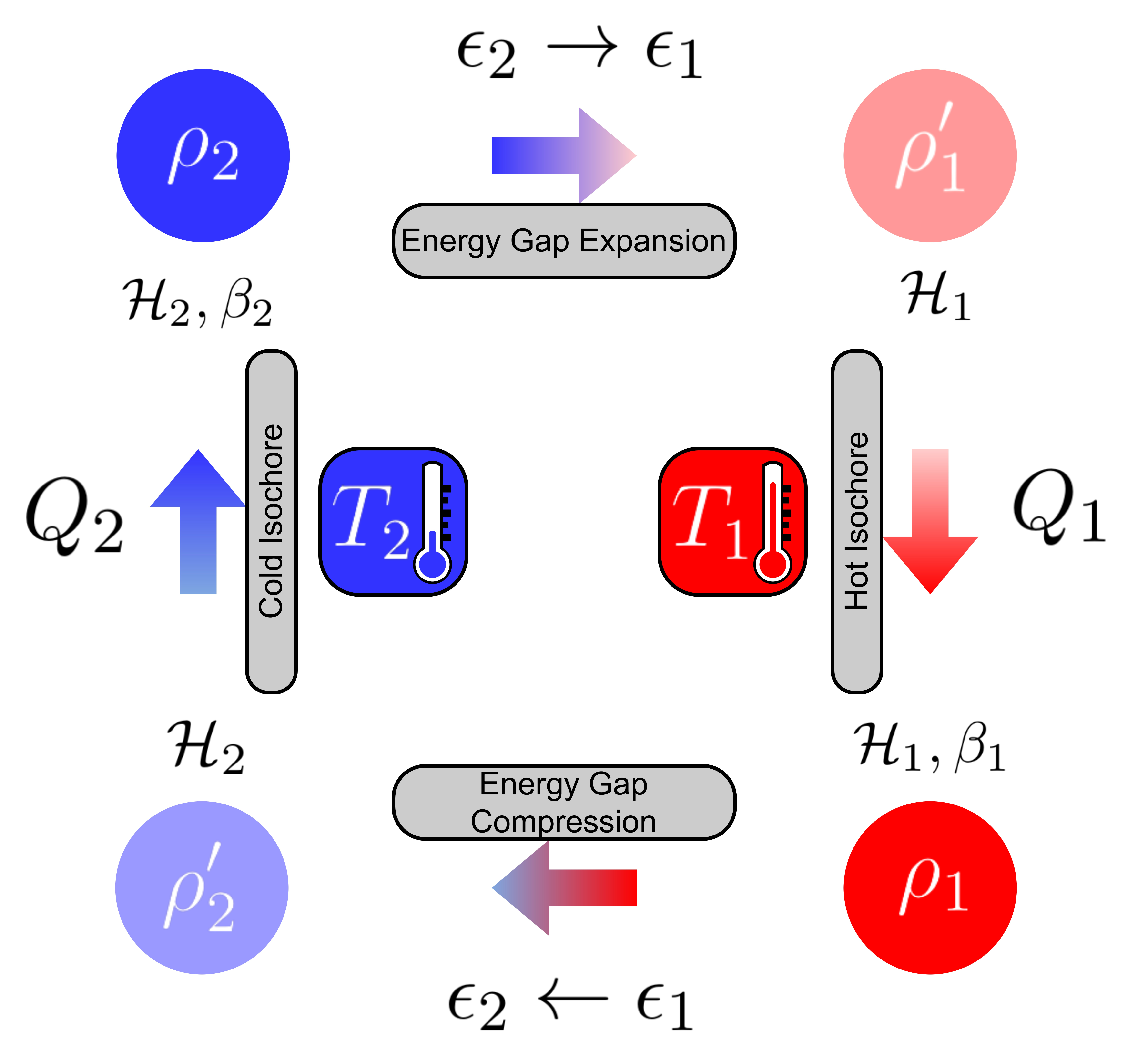}
	\caption{The schematic representation of quantum Otto cycle with a single qubit}
	\label{OttoCycle}
\end{figure}
\begin{figure}[tb]
	\centering
	\includegraphics[scale=0.7]{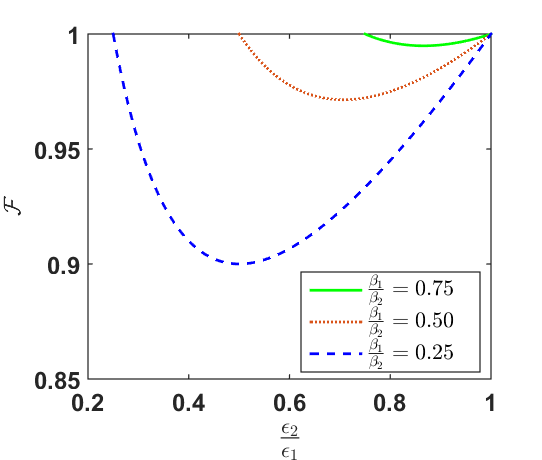}
	\caption{The fidelity, $\mathcal{F}$ as a function of the ratio $\frac{\epsilon_{2}}{\epsilon_{1}}$ is plotted for $W=\eta=0$ at different inverse temperature ratios $\frac{\beta_{1}}{\beta_{2}}$. The value of $\beta_{1}\epsilon_{1}=0.01$ is fixed for all the cases.}
	\label{Fplot}
\end{figure}
An ideal Quantum Otto Cycle (QOC) is the quantum-mechanical analog of the classical Otto cycle, comprising two isochoric and two adiabatic processes. The quantum isochoric process is characterized by the invariance of the energy eigenvalues throughout the process, while the quantum adiabatic process demands not only thermal isolation but also the preservation of the probability of occupation associated with the eigenstates~\cite{QuanQuantum2007}. We consider a single qubit as the working medium and operate it through a non-adiabatic Otto cycle depicted in Figure~\ref{OttoCycle}. The finite-time unitary processes replace the adiabatic processes to construct the non-adiabatic Otto cycle. An outline of this cyclic process as shown in Figure~\ref{OttoCycle} is given as follows.\\
\textbf{Hot isochore (HI):} The qubit under a constant Hamiltonian $\mathcal{\hat{H}}_{1}$ may thermally equilibrate with the hot bath at the temperature $T_{1}$. The resultant thermal state is $\rho_{1}$.\\ 
\textbf{Energy gap compression (EGC):} Following the initial thermalization, we isolate the qubit from the hot bath and drive it using a time-varying Hamiltonian $\mathcal{\hat{H}}(t)$. The Hamiltonian changes from $\mathcal{\hat{H}}\left(0\right)=\mathcal{\hat{H}}_{1}$ to $\mathcal{\hat{H}}\left(\tau\right)=\mathcal{\hat{H}}_{2}$ and the energy gap between the eigenstates compresses during the time period $\tau$. The resultant qubit state after this finite time evolution is $\rho^{\prime}_{2}$.\\
\textbf{Cold isochore (CI):} The qubit under the constant Hamiltonian $\mathcal{\hat{H}}_{2}$ is attached to a cold bath at temperature $T_{2}$ until thermalization. The thermalized state after CI is $\rho_{2}$.\\
\textbf{Energy gap expansion (EGE):} Detach the qubit from cold bath and drive it using $\mathcal{\hat{H}}\left(\tau-t\right)$ back to $\mathcal{\hat{H}}_{1}$ from $\mathcal{\hat{H}}_{2}$. We assume, EGE takes the same time duration $\tau$ of the compression process as a result of time reversal of the Hamiltonian. The energy gap between the eigenstates expands and the evolved qubit state is $\rho^{\prime}_{1}$.

In general, it is possible to map the dynamics of any particular qubit to the generic spin Hamiltonian,
\begin{equation}
	\mathcal{\hat{H}}(t)=X\sigma_{x}+Y\sigma_{y}+Z\sigma_{z},
	\label{GHamil}
\end{equation}
with $\sigma_{x},\sigma_{y}$ and $\sigma_{z}$ are the Pauli matrices, and $X,Y$ and $Z$ are the time-dependent control parameters of the qubit. The absolute energy eigenvalue of the above Hamiltonian is $\epsilon=\vert\pm\sqrt{X^{2}+Y^{2}+Z^{2}}\vert$ and the eigenstates at a particular instant of time are
\begin{equation}
	\vert \Phi^{1}\rangle=\begin{bmatrix}
		\sqrt{\frac{\left(X+iY\right)\left(X-iY\right)}{2\epsilon\left(\epsilon+Z\right)}}\\
		-\sqrt{\frac{\left(\epsilon+Z\right)\left(X+iY\right)}{2\epsilon\left(X-iY\right)}}
	\end{bmatrix},~~~~~\vert\Phi^{2}\rangle=\begin{bmatrix}
		\sqrt{\frac{\left(X+iY\right)\left(X-iY\right)}{2\epsilon\left(\epsilon-Z\right)}}\\
		\sqrt{\frac{\left(\epsilon-Z\right)\left(X+iY\right)}{2\epsilon\left(X-iY\right)}}
	\end{bmatrix}.
	\label{epsilon}
\end{equation}
We can assume that the Hamiltonian $\mathcal{\hat{H}}_{1}$ has energy eigenvalues $- \epsilon_{1}$ and $+ \epsilon_{1}$ corresponding to the eigenstates $\vert \Phi_{1}^{1}\rangle$ and $\vert \Phi_{1}^{2}\rangle$ respectively. Similarly, the Hamiltonian $\mathcal{\hat{H}}_{2}$ has eigenvalues $-\epsilon_{2}$ and $+ \epsilon_{2}$ correspondig to the eigenstates $\vert \Phi_{2}^{1}\rangle$ and $\vert \Phi_{2}^{2}\rangle$ respectively. For an adiabatically driven energy gap compression process, the physical state of the qubit shows same probability of occupancy among the eigenstates of $\mathcal{\hat{H}}_{1}$ and among the eigenstates of $\mathcal{\hat{H}}_{2}$ respectively before and after the process. Similarly, the probability of occupation among the eigenstates of $\mathcal{\hat{H}}_{2}$ do not change as the Hamiltonian slowly changes to $\mathcal{\hat{H}}_{1}$ during the EGE process. However, the finite-time processes, EGC and EGE with time-varying Hamiltonian generates transition probability among the eigenstates, and this will cause the quantum non-adiabaticity in the cycle~\cite{KatoOn1950}. In other words, the isolation from the thermal baths (hot/cold) is not sufficient for a quantum adiabatic process, but it needs a prolonged variation of the Hamiltonian with time~\cite{BornBeweis1928}. The finite-time dynamics of these processes may not preserve the probability of occupation associated with the eigenstates. Based on the formulation of quantum mechanics, the finite-time evolution operator of the EGC process is given by~\cite{SakuraiModern1994},
\begin{equation}
	U_{egc}=\mathcal{T}exp\left({-i\int_{0}^{\tau}\mathcal{\hat{H}}(t) dt}\right).
	\label{evolution}
\end{equation}
Similarly, we can define the time evolution operator for EGE process as
\begin{equation}
U_{ege}=\mathcal{T}exp\left({-i\int_{0}^{\tau}\mathcal{\hat{H}}(\tau-t) dt}\right).
\label{Unitaryoperator}
\end{equation}
The $\mathcal{T}exp\left(\right)$ represents the time ordered exponential. The time ordering is very important with qubit, as in general, the time-dependent Hamiltonian at an instant may not commute with the Hamiltonian at another instant of time. The above time evolution operators can be constructed by analytical or numerical methods~\cite{SchmidtkeStiffness2018,FeiguinTime2005}. In this paper, we have followed a numerical method (see Appendix A for details). 

The density matrix of the qubit after each thermalization is equivalent to the quantum counterpart of the canonical ensemble~\cite{VinjanampathyQuantum2016}. The density matrix of the qubit following the HI process is $\rho_{1}=\frac{e^{-\beta_{1}\mathcal{\hat{H}}_{1}}}{\mathcal{Z}_{1}}$ and the density matrix after the CI process is $\rho_{2}=\frac{e^{-\beta_{2}\mathcal{\hat{H}}_{2}}}{\mathcal{Z}_{2}}$, where the inverse temperature $\beta_{1}=\frac{1}{k_{B}T_{1}}$ $\left(\beta_{2}=\frac{1}{k_{B}T_{2}}\right)$ and the partition function $\mathcal{Z}_{1}$ $\left(\mathcal{Z}_{2}\right)$ corresponding to the hot (cold) bath with temperature $T_{1}$ ($T_{2}$). The term, $k_{B}$ is the Boltzmann constant. The density matrix of the evolved qubit during the EGC process is
\begin{equation}
	\rho^{\prime}_{2}=U_{egc}\rho_{1}U_{egc}^{\dagger}.
	\label{DenMatEvo1}
\end{equation}
Similarly, the density matrix following the EGE process is
\begin{equation}
	\rho^{\prime}_{1}=U_{ege}\rho_{2}U_{ege}^{\dagger}.
	\label{DenMatEvo2}
\end{equation}
 The heat exchanged $Q_{1}$ $\left(Q_{2}\right)$ during hot (cold) isochore characterizes the thermodynamics of the cycle. The standard definition of heat exchanged is $Q=\int_{0}^{\tau}trace\left(\dot{\rho}H(t)\right)dt$, where $\rho$ is the time-dependent density matrix of the system~\cite{VinjanampathyQuantum2016}. However, the HI (CI) under constant Hamiltonian produce zero work~\cite{SolfanelliNonadiabatic2020,SuThe2018} and gives $Q_{1}$ $\left(Q_{2}\right)$ as the difference in expectation value of the energy before and after the thermalization step. The thermal expectation value of energy after the HI (CI) process is $trace\left\{\mathcal{\hat{H}}_{1}\rho_{1}\right\}$ $\left(trace\left\{\mathcal{\hat{H}}_{2}\rho_{2}\right\}\right)$. Similarly, the expectation value of energy before HI (CI) is $trace\left\{\mathcal{\hat{H}}_{1}\rho^{\prime}_{1}\right\}$ $\left(trace\left\{\mathcal{\hat{H}}_{2}\rho^{\prime}_{2}\right\}\right)$. Calculating the above traces and subtracting the expectation value of energy before the thermalization process from the expectation value of energy after the thermalization process gives~\cite{SolfanelliNonadiabatic2020}
\begin{equation}
	Q_{1}=-\epsilon_{1}tanh(\beta_{1}\epsilon_{1})-\epsilon_{2}~tanh\left(\beta_{1}\epsilon_{1}\right)\left(1-2\mathcal{F}_{1}\right),
	\label{Q1}
\end{equation}
\begin{equation}
	Q_{2}=-\epsilon_{2}tanh(\beta_{2}\epsilon_{2})-\epsilon_{1}~tanh\left(\beta_{2}\epsilon_{2}\right)\left(1-2\mathcal{F}_{2}\right),
	\label{Q2}
\end{equation}
where $\mathcal{F}_{1}$ and $\mathcal{F}_{2}$ are fidelities defined as
\begin{equation}
	\mathcal{F}_{1}=\vert\langle\Phi^{i}_{1}\vert U_{ege}\vert\Phi^{i}_{2}\rangle\vert^{2}, 
	\label{fidelity2}
\end{equation} 
\begin{equation}
	\mathcal{F}_{2}=\vert\langle\Phi^{i}_{2}\vert U_{egc}\vert\Phi^{i}_{1}\rangle\vert^{2}. 
	\label{fidelity1}
\end{equation}
Fidelity is an inner product relation between two normalized quantum states with values ranging from zero to one. An evolved state, which is completely distinguishable from (orthogonal to) the corresponding adiabatic final state gives zero fidelity, while the unit fidelity corresponds to the adiabatic evolution of the qubit (evolved state is the same as the adiabatic final state). Thus, a fidelity less than unity measures the deviation of the evolved state from the adiabatic final state of the qubit due to the finite-time dynamics. It is worth noticing that we require the fidelities for the eigenstates of the Hamiltonian $\mathcal{\hat{H}}_{1}$ and $\mathcal{\hat{H}}_{2}$ to calculate the heat exchanged, while the actual EGC and EGE processes start with a mixed state of the qubit thermalized with the corresponding bath. We can use equation (\ref{epsilon}) to obtain the specific form of the eigenstates $\vert\Phi^{i}_{1}\rangle$ and $\vert\Phi^{i}_{2}\rangle$. It is possible to choose either the ground state ($i=1$) or the excited state ($i=2$) to calculate the fidelity, since the value of $\mathcal{F}_{1}$ and $\mathcal{F}_{2}$ depend only on the unitary operators $U_{ege}$ and $U_{egc}$. In this paper, we work with the Hamiltonian $\mathcal{\hat{H}}(t)$, which is invariant under the application of the time-reversal operator at any instant of time $0\leq t\leq \tau$. As the EGE is executed by reversing the time parameter of EGC Hamiltonian, the above condition results in the same fidelity for both the unitary operations ($\mathcal{F}_{1}=\mathcal{F}_{2}=\mathcal{F}$)~\cite{SolfanelliNonadiabatic2020}.

The work done by a quantum system is defined as $W=\int_{0}^{\tau}trace\left(\rho\dot{H}(t)\right)dt$~\cite{VinjanampathyQuantum2016}. Since the change in internal energy of the engine during a complete cycle is zero, the total work performed by the qubit is the sum of the heat exchanged during the isochoric processes, i.e.,
\begin{equation}
	W=Q_{1}+Q_{2}.
	\label{W}
\end{equation}
A working QHE is possible, if the total work ($W$) obtained and heat ($Q_{1}$) absorbed from the hot bath are positive for $0<\beta_{1}<\beta_{2}$~\cite{SolfanelliNonadiabatic2020}, such that it should satisfy the Clausius inequality~\cite{FermiThermodynamics1956},
\begin{equation}
	Q_{1}\beta_{1}+Q_{2}\beta_{2}\leq 0.
\label{CE}
\end{equation}
As a natural consequence, the efficiency ($\eta=\frac{W}{Q_{1}}$) will also be positive. The expanded expression for the total work can be obtained from equations (\ref{Q1})-(\ref{W}) as
\begin{equation}
	W=-tanh\left(\beta_{1}\epsilon_{1}\right)\left[\epsilon_{1}+\epsilon_{2}\left(1-2\mathcal{F}\right)\right]-tanh\left(\beta_{2}\epsilon_{2}\right)\left[\epsilon_{2}+\epsilon_{1}\left(1-2\mathcal{F}\right)\right].
	\label{workdone}
\end{equation}
From the above equation we can find the fidelity $\mathcal{F}$ for $W=\eta=0$ as
\begin{equation}
	\mathcal{F}=\frac{1}{2}\left(1+\frac{\epsilon_{1}tanh\left(\beta_{1}\epsilon_{1}\right)+\epsilon_{2}tanh\left(\beta_{2}\epsilon_{2}\right)}{\epsilon_{2}tanh\left(\beta_{1}\epsilon_{1}\right)+\epsilon_{1}tanh\left(\beta_{2}\epsilon_{2}\right)}\right),
	\label{fidelityequation}
\end{equation}
 and plot it as a function of $\frac{\epsilon_{2}}{\epsilon_{1}}$ as shown in Figure \ref{Fplot} for different values of temperature ratio $\frac{\beta_{1}}{\beta_{2}}$ by keeping $\beta_{1}\epsilon_{1}$ as a constant. The total work and efficiency are zero throughout the plotted lines. Thus, crossing a line from one side to another will change the sign of total work and efficiency. This change in the sign allows us to analyze the values of fidelity for positive total work at a particular ratio of energy eigenvalues ($\frac{\epsilon_{2}}{\epsilon_{1}}$) corresponding to a working QHE. The fidelity below the line corresponds to the negative work output. The directional compatibility of heat exchanged, i.e., the sign of $Q_{1}$ and $Q_{2}$, according to equations (\ref{W}) and (\ref{CE}), restricts the Otto cycle to function as a heater ($Q_{1}\leq 0,Q_{2}\leq 0$), accelerator ($Q_{1}\geq 0,Q_{2}\leq 0$) and Refrigerator ($Q_{1}\leq 0,Q_{2}\geq 0$) with negative work output. The upper region gives the fidelity corresponding to the positive work output and exclusively allows the construction of a heat engine ($Q_{1}\geq 0,Q_{2}\leq 0$)~\cite{SolfanelliNonadiabatic2020}. The broken line plot (Figure \ref{Fplot}) for $\frac{\beta_{1}}{\beta_{2}}=0.25$ shows more range of fidelity, $\mathcal{F}$ with positive work output and the range of fidelity decreases with increasing $\frac{\beta_{1}}{\beta_{2}}$. Also, the area covered by the lines shrinks towards $\mathcal{F}=1$ for higher values of $\frac{\beta_{1}}{\beta_{2}}$. This indicates that if the temperature gradient between the hot and cold baths is small, we need high fidelity processes after each thermalization to achieve positive work. It satisfies the intuition that there is more chance to get work output from the heat engine by finite time drive if the temperature of the hot bath is much greater than that of the cold bath. In other words, we can expect at least some work output after deduction of all frictional loss if there is enough freedom to change the system energy and that freedom increases with the temperature gradient between the baths. The maximum work output is always bounded by the quasi-static ($\mathcal{F}=1$) case. However, giving more temperature gradient allows finite-time driving (corresponding to $\mathcal{F}<1$) compromising on the work output. The ratio $\frac{\epsilon_{2}}{\epsilon_{1}}$ for a working QHE varies from $\frac{\beta_{1}}{\beta_{2}}$ to 1, implying $\epsilon_{2}<\epsilon_{1}$. Thus, the energy gap between the eigenvalues reduces during the EGC process and it increases to the earlier value during the EGE process. The required fidelity at $\frac{\epsilon_{2}}{\epsilon_{1}}=\frac{\beta_{1}}{\beta_{2}}$ and $\frac{\epsilon_{2}}{\epsilon_{1}}=1$ is unity and it results in zero total work. We can design a QHE with a positive work output and efficiency by understanding all the above conditions. 
 
 We have to calculate the cost for generating the fields associated with the Hamiltonian, accounting the use of resources for the cyclic operation of the QHE. There are several ways to calculate it~\cite{CampoShortcuts2019}, but we follow the one explained in the Refs.~\cite{ZhengCost2016,AbahEnergetic2019}. The formulation of cost depends on the control mechanism involved in the implementation of the QHE. We assume implementing the Hamiltonian for QHE using electromagnetic waves in all the subsequent sections. The power required to generate these electromagnetic waves will scale as the square of electric fields associated with it, and the cost for implementation~\cite{ZhengCost2016},
 \begin{equation}
 	\mathcal{C}\propto \int_{0}^{\tau} \vert\vert \mathcal{\hat{H}}(t)\vert\vert^{2}dt,
 	\label{cost}
 \end{equation} 
where, the Frobenius norm $\vert\vert \mathcal{\hat{H}}(t)\vert\vert=\sqrt{trace\left(\mathcal{\hat{H}}^{\dagger}\mathcal{\hat{H}}\right)}$ and $\tau$ is the time duration of the process.

In this section, we have defined all the parameters to understand the performance of a QHE. The following section deliver a particular case of Landau-Zener Hamiltonian $\mathcal{\hat{H}}_{LZ}(t)$, and check for the possibility of a working non-adiabatic QHE. We will restrict the study to the inverse temperature ratio $\frac{\beta_{1}}{\beta_{2}}=0.25$ in all further analysis as it widens the spectrum of allowed fidelity values for positive total work; thus, the possibility of a non-adiabatic QHE.
\section{Qubit heat engine with Landau-Zener dynamics}
\label{QHELZ}
\begin{figure}[tb]
	\centering
	\begin{subfigure}{.45\linewidth}
		\includegraphics[scale=0.6]{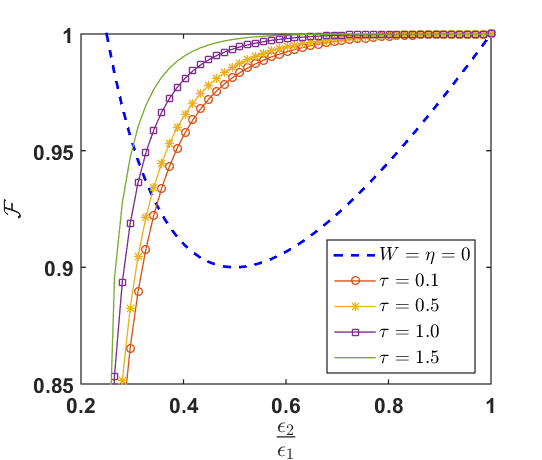}
		\subcaption{}
		\label{PlotForLZNormal}
	\end{subfigure}
	\begin{subfigure}{.45\linewidth}
		\includegraphics[scale=0.6]{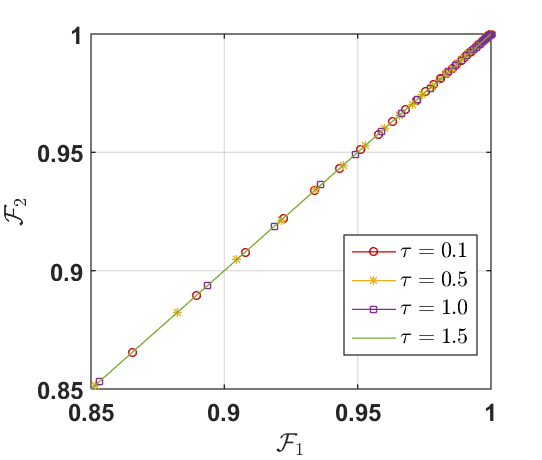}
		\subcaption{}
		\label{F1F2Plot}
	\end{subfigure}	
	\caption{Plot (a) shows the fidelity $\mathcal{F}$ of EGC and EGE using LZ Hamiltonian as a function of $\frac{\epsilon_{2}}{\epsilon_{1}}$ for different time durations $\tau$ of the process. The fidelity plot for $W=\eta=0$ is shown as a reference to understand the required fidelity for positive work output and positive efficiency. Plot (b) shows the equivalence of fidelity of EGE process, $\mathcal{F}_{1}$ and the fidelity of EGC process, $\mathcal{F}_{2}$ using the numerical approach provided in Appendix A. The temperature gradient $\frac{\beta_{1}}{\beta_{2}}$ is fixed as 0.25. Also, we fixed the value of $\beta_{1}\epsilon_{1}=0.01$ and changed $\epsilon_{2}$ for plotting the figure.}
	\label{LZFplots}
\end{figure}
The Landau-Zener (LZ) model is proposed initially as a solvable model for a two-level system with coupled eigenstates and linear variation of energy eigenvalue~\cite{LandauZur1932}. We can understand the non-adiabatic effects of LZ dynamics from the expression derived in Refs.~\cite{LandauZur1932,ZenerNon1932}. In literature, the exploration of LZ dynamics extended to systems governed by non-linear equations~\cite{ArturLandau2004}, non-linear time-dependence of energy eigenvalues~\cite{VitanovNon1999} and also with non-Hermitian Hamiltonian~\cite{BoyanPseudo2017}. In this work, we stick on to the original LZ Hamiltonian,
\begin{equation}
	\mathcal{\hat{H}}_{LZ}(t)=Z\sigma_{z}+X\sigma_{x},
	\label{LZHamil}
\end{equation} 
describes the dynamics of the two-level system for the linear modulation of $Z$ with time, through an avoided level crossing with an energy gap of $2X$, where $X$ is a constant and $\sigma_{x},\sigma_{z}$ are Pauli spin operators~\cite{ZhengCost2016,CakmakSpin2019,LandauZur1932,ZenerNon1932}. On the basis with $\sigma_{z}$ and $\sigma_{x}$ are real, the absolute energy eigenvalue of the above Hamiltonian is
\begin{equation}
	\epsilon=\vert\pm \sqrt{Z^{2}+X^{2}}\vert=\sqrt{Z^{2}+X^{2}},
	\label{epsilonkl}
\end{equation}
and the normalized orthogonal eigenstates are,
\begin{equation}
	\vert \Phi^{1}\rangle=\begin{bmatrix}
	\sqrt{\frac{X^{2}}{2\epsilon\left(\epsilon+Z\right)}}\\
	-\sqrt{\frac{\left(\epsilon+Z\right)}{2\epsilon}}
	\end{bmatrix},
	\label{EigStatOfHLz1}
\end{equation}
\begin{equation}
\vert\Phi^{2}\rangle=\begin{bmatrix}
\sqrt{\frac{X^{2}}{2\epsilon\left(\epsilon-Z\right)}}\\
\sqrt{\frac{\left(\epsilon-Z\right)}{2\epsilon}}
\end{bmatrix}.
\label{EigStatOfHLz2}
\end{equation}
We consider a real, linear function `Z', which varies from $z_{1}$ to $z_{2}$ as
\begin{equation}
	Z=z_{1}+\Delta_{z}\frac{t}{\tau},
	\label{Zeq}
\end{equation}
where $\Delta_{z}=z_{2}-z_{1}$ and the constant,
\begin{equation}
 	X=x.
 	\label{Xeq}
\end{equation}
We use the time reversed protocol explained in section two, that the Hamiltonian $\mathcal{\hat{H}}_{LZ}(t)$ changes from $\mathcal{\hat{H}}_{LZ}(0)=\mathcal{\hat{H}}_{1}$ to $\mathcal{\hat{H}}_{LZ}(\tau)=\mathcal{\hat{H}}_{2}$ during the EGC process with $0\leq t\leq \tau$, while the EGE process under the Hamiltonian $\mathcal{\hat{H}}_{LZ}(\tau-t)$ changes from $\mathcal{\hat{H}}_{2}$ to $\mathcal{\hat{H}}_{1}$ with $0\leq t\leq \tau$. As the initial value of the function $Z=z_{1}$, the absolute energy eigenvalue of $\mathcal{\hat{H}}_{1}$ is $\epsilon_{1}=\sqrt{z_{1}^{2}+x^{2}}$, and the corresponding eigenstates are 
 \begin{equation}
 	\vert \Phi^{1}_{1}\rangle=\begin{bmatrix}
 		\sqrt{\frac{x^{2}}{2\epsilon_{1}\left(\epsilon_{1}+z_{1}\right)}}\\
 		-\sqrt{\frac{\left(\epsilon_{1}+z_{1}\right)}{2\epsilon_{1}}}
 	\end{bmatrix},~~~~~~~~~~\vert\Phi^{2}_{1}\rangle=\begin{bmatrix}
 	\sqrt{\frac{x^{2}}{2\epsilon_{1}\left(\epsilon_{1}-z_{1}\right)}}\\
 	\sqrt{\frac{\left(\epsilon_{1}-z_{1}\right)}{2\epsilon_{1}}}
 \end{bmatrix}.
\end{equation}
Similarly, we can find the eigenvalue ($\epsilon_{2}$) and eigenstates ($\vert\Phi^{1}_{2}\rangle$ and $\vert\Phi^{2}_{2}\rangle$) corresponding to $\mathcal{\hat{H}}_{2}$ from equations (\ref{epsilonkl}),(\ref{EigStatOfHLz1}) and (\ref{EigStatOfHLz2}) for the final value of $Z=z_{2}$. The energy gap at any stage of the cycle ($2\epsilon_{1}$ or $2\epsilon_{2}$) is a finite positive value. Also, considering the fact that the energy gap reduces during EGC ($\epsilon_{2}<\epsilon_{1}$), we fix the value of $\epsilon_{1}$ in such a way that the value of $\epsilon_{2}\ge 0$ for all possible values of $\left(\frac{\epsilon_{2}}{\epsilon_{1}}\right)$. Thus, we construct the condition for $z_{1}$ from the expressions for $\epsilon_{1}$ and $\epsilon_{2}$ as
 \begin{equation}
 	z_{1}^{2}\ge  \frac{x^{2}\left(1-\left(\frac{\epsilon_{2}}{\epsilon_{1}}\right)^{2}\right)}{\left(\frac{\epsilon_{2}}{\epsilon_{1}}\right)^{2}}.
 	\label{f1ofzero}
 \end{equation}
 \begin{figure}[tb]
 	\centering
 	\begin{subfigure}{.49\linewidth}
 		\includegraphics[scale=0.7]{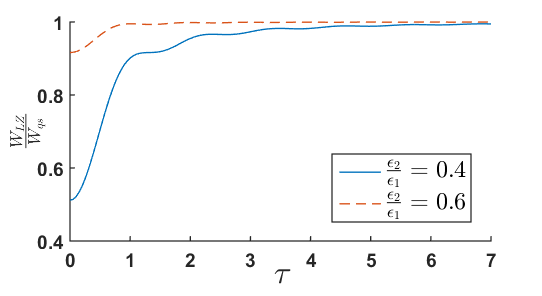}
 		\subcaption{}
 		\label{WorkPlotForLZ}
 	\end{subfigure}
 	\begin{subfigure}{.49\linewidth}
 		\includegraphics[scale=0.7]{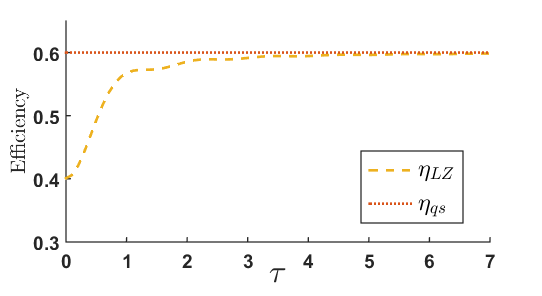}
 		\subcaption{}
 		\label{EffcLZSH4}
 	\end{subfigure}
 	\begin{subfigure}{.49\linewidth}
 		\includegraphics[scale=0.7]{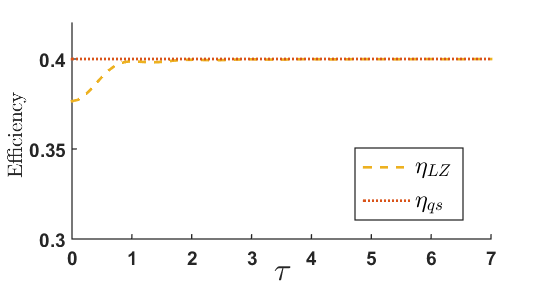}
 		\subcaption{}
 		\label{EffcLZSH6}
 	\end{subfigure}	
 	\caption{The plot (a) shows the ratio of $W_{LZ}$, the work from the LZ Hamiltonian based QHE to $W_{qs}$, the quasi-static work against the time duration $\tau$ of the finite-time unitary processes. The plot for the efficiency of the LZ Hamiltonian based QHE versus the time duration $\tau$ of the finite-time unitary processes for the ratio of energy eigenvalue, $\left(b\right)$ $\frac{\epsilon_{2}}{\epsilon_{1}}=0.4$ and $\left(c\right)$ $\frac{\epsilon_{2}}{\epsilon_{1}}=0.6$. The corresponding quasi-static efficiency $\left(\eta_{qs}\right)$ is also included in each plot. All the other parameters are fixed as that of the Figure \ref{LZFplots}.}
 	\label{EffcLZSH}
 \end{figure}
\begin{figure}[b]
	\centering
	\begin{subfigure}{.49\linewidth}
		\includegraphics[scale=.7]{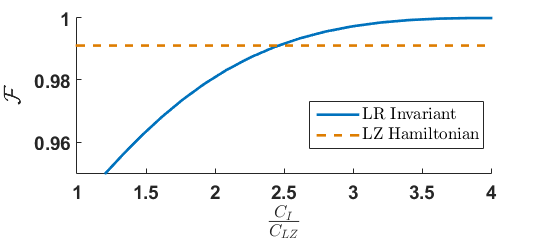}
		\subcaption{}
		\label{FidalityInvariant04}
	\end{subfigure}
	\begin{subfigure}{.49\linewidth}
		\includegraphics[scale=.7]{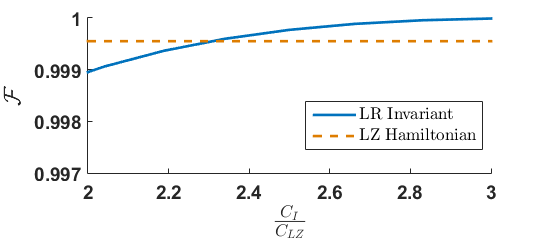}
		\subcaption{}
		\label{FidalityInvariant06}
	\end{subfigure}
	
	\caption{Fidelity of the unitary processes using LR invariant is plotted against the cost ratio, $\frac{C_{I}}{C_{LZ}}$ for time duration, $\tau=1.0$. The corresponding fidelity of the QHE with LZ Hamiltonian is included in each plot for comparison. The change in the energy gap is fixed corresponding to the ratio, (a)$\frac{\epsilon_{2}}{\epsilon_{1}}=0.4$ and (b)$\frac{\epsilon_{2}}{\epsilon_{1}}=0.6$. $C_{I}$ is changed by the parametric variation of $x(t),y(t)$ and $z(t)$ with constant $X$ given by equations (\ref{xt}),(\ref{yt}),(\ref{zt}) and (\ref{Xeq}) respectively.}
	\label{FI}
\end{figure}
 We executed both EGC and EGE computationally using the formulas derived in section~\ref{NonAQOC} (Eq. (\ref{evolution}) to Eq. (\ref{DenMatEvo2})) by assuming that the qubit is thermalized to the lowest energy eigenstate $\vert\phi_{1}^{1}\rangle$ of $\mathcal{\hat{H}}_{1}$ during HI, and $\vert\phi_{2}^{1}\rangle$ of $\mathcal{\hat{H}}_{2}$ during CI. The fidelity ($\mathcal{F}$) plots for EGC and EGE with possible values of $\frac{\epsilon_{2}}{\epsilon_{1}}$ at various time durations $\tau$ is plotted in Figure \ref{PlotForLZNormal}. The fidelity plot corresponding to $W=\eta=0$ (Figure \ref{Fplot}) is superimposed with Figure \ref{PlotForLZNormal} to understand the values of $\mathcal{F}$ resulting in positive work output and positive efficiency. Increasing the time duration includes more, larger possible changes in the energy gap with the positive work output. The Hamiltonian in Eq. (\ref{LZHamil}) becomes time-independent for $\frac{\epsilon_{2}}{\epsilon_{1}}=1$ ($Z$ will be constant) and cause to result exactly in its eigenstates (satisfies the time-independent Schrodinger equation) irrespective of the time duration of the process. Thus, the fidelity is always 1 for such processes, and it is evident from the plots that $\mathcal{F}=1$ at $\frac{\epsilon_{2}}{\epsilon_{1}}=1$ for all the time durations. We have compiled both EGC and EGE separately, and the obtain $\mathcal{F}_{1}$ and $\mathcal{F}_{2}$ as given in the Figure \ref{F1F2Plot}, which shows that our numerical approach is highly dependable as it satisfies the condition $\mathcal{F}_{1}=\mathcal{F}_{2}=\mathcal{F}$ for all the plotted time durations.
 
 The work $W_{LZ}$ and the heat absorbed from the hot bath can be calculated using the equations (\ref{W}) and (\ref{Q1}) respectively, by substituting the corresponding fidelity obtained from the dynamics of the LZ Hamiltonian. The work ratio $\frac{W_{LZ}}{W_{qs}}$ is plotted in Figure \ref{WorkPlotForLZ}, where $W_{qs}$ is the quasi-static work output obtained using equation~\cite{QuanQuantum2007}, $$W_{qs}=2\left(\epsilon_{1}-\epsilon_{2}\right)\left(\frac{1}{1+exp\left(2\beta_{1}\epsilon_{1}\right)}-\frac{1}{1+exp\left(2\beta_{2}\epsilon_{2}\right)}\right).$$ As we expected, the work obtained from the QHE with LZ dynamics deviates from the quasi-static value for the lower time durations. Also, the non-adiabaticity is more evident for larger change in the energy gap (smaller the ratio $\frac{\epsilon_{2}}{\epsilon_{1}}$) during EGC and EGE. The efficiency, $\eta_{LZ}$ obtained using $W_{LZ}$ and corresponding $Q_{1}$ is plotted in Figure \ref{EffcLZSH4} for $\frac{\epsilon_{2}}{\epsilon_{1}}=0.4$, and Figure \ref{EffcLZSH6} for $\frac{\epsilon_{2}}{\epsilon_{1}}=0.6$. The efficiency converges to the corresponding quasi-static value~\cite{QuanQuantum2007}, $$\eta_{qs}=1-\frac{\epsilon_{2}}{\epsilon_{1}}$$ at large time durations. The non-adiabatic evolution cause the decrease in efficiency from the quasi-static value for QHE involving faster EGC and EGE processes. All the above observations conclude that the implications of non-adiabaticity are more profound in lower time duration as the engine asymptotically approaches the adiabatic performance at a large time durations. Using the STA techniques, we can improve the engine's performance at a lower time duration up to that of the adiabatic one. In the following section, we use the LR invariant for the LZ Hamiltonian to enhance the performance of the QHE. We discuss how it is implemented and the relative cost in improving the performance of the engine. 
\section{Shortcut to adiabaticity using LR invariant}
\label{STAI}
The performance of LZ Hamiltonian based QHE shows deviation from the quasi-static performance of the engine at lower time durations. The non-adiabaticity in the states evolved under the LZ dynamics can be steered and get to an evolved state having unit fidelity using an LR invariant~\cite{LewisAn1969} of $\mathcal{\hat{H}}_{LZ}$. In general, an invariant $\mathcal{\hat{I}}$ and the Hamiltonian $\mathcal{\hat{H}}$ are related to each other by $\frac{\partial\mathcal{\hat{I}}}{\partial t}=i\left[\mathcal{\hat{I}},\mathcal{\hat{H}}\right]$, where $\left[~,~\right]$ is the commutator. Usually, we derive the invariant corresponding to a Hamiltonian and inverse engineer the Hamiltonian control parameters. This inverse engineering is carried out by fixing some boundary conditions for adiabatic final states. Further, realizing the inverse engineered Hamiltonian allows us to drive the system in a required STA path. Another possibility is to drive the system described by the invariant itself to achieve STA of the corresponding Hamiltonian. We have to construct a Hermitian invariant to realize both of the above methods collectively called STA using the LR invariant. The LR invariant corresponding to the LZ Hamiltonian,
\begin{figure}[t]
	\centering
	\begin{subfigure}{.49\linewidth}
		\includegraphics[scale=.55]{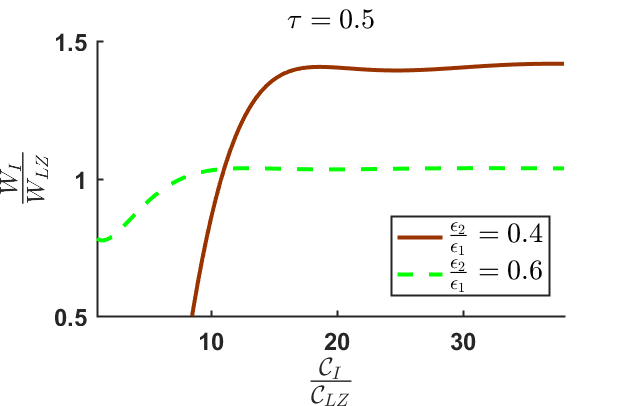}
		\subcaption{}
		\label{WRatio5}
	\end{subfigure}
	\begin{subfigure}{.49\linewidth}
		\includegraphics[scale=.55]{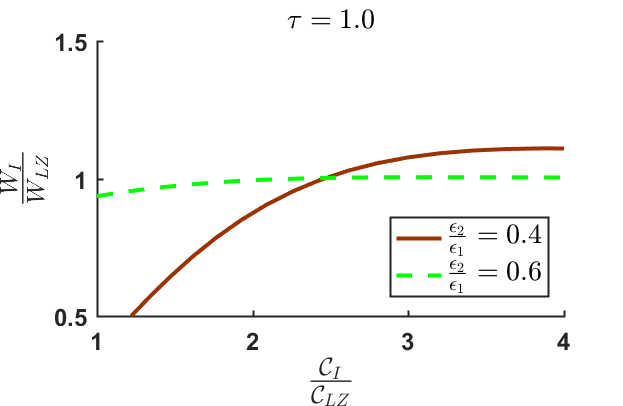}
		\subcaption{}
		\label{WRatio1}
	\end{subfigure}
	\begin{subfigure}{.49\linewidth}
		\includegraphics[scale=.55]{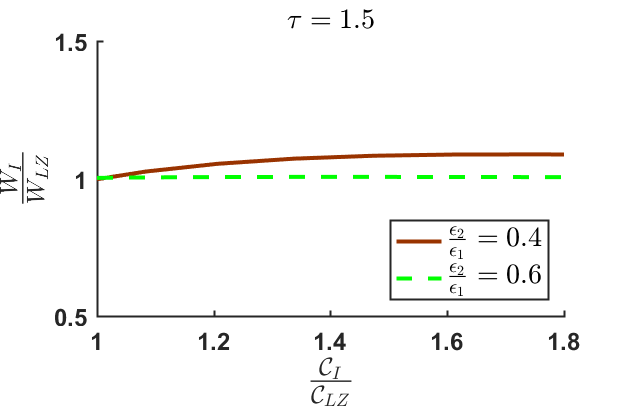}
		\subcaption{}
		\label{WRatio15}
	\end{subfigure}
	\caption{The work ratio, $\frac{W_{I}}{W_{LZ}}$ is plotted against the ratio of costs, $\frac{\mathcal{C}_{I}}{\mathcal{C}_{LZ}}$ at different values of $\frac{\epsilon_{2}}{\epsilon_{1}}$ for the time duration $\left(a\right)$ $\tau=0.5$, $\left(b\right)$ $\tau=1.0$ and $\left(c\right)$ $\tau=1.5$. $C_{I}$ is changed by the parametric variation of $x(t),y(t)$ and $z(t)$ with constant $X$ given by equations (\ref{xt}),(\ref{yt}),(\ref{zt}) and (\ref{Xeq}) respectively.}
	\label{WR}
\end{figure}
$\mathcal{\hat{H}}_{LZ}$~(\ref{LZHamil}) is assumed to be of the form,
\begin{equation}
	\mathcal{\hat{I}}=x(t)\sigma_{x}+y(t)\sigma_{y}+z(t)\sigma_{z},
	\label{Invariantform}		
\end{equation}
with control parameters $x(t),y(t),z(t)$ in all the three mutually perpendicular axes. The above particular form of invariant satisfies the condition, $\frac{\partial \mathcal{\hat{I}}}{\partial t}=i\left[\mathcal{\hat{I}},\mathcal{\hat{H}}_{LZ}\right]$ to stay conserved in the Hilbert space of $\mathcal{\hat{H}}_{LZ}$. Substituting for the Hamiltonian $\mathcal{\hat{H}}_{LZ}$ and the invariant $\mathcal{\hat{I}}$ in the above commutation relation, we obtain a mutually dependent set of conditions,
\begin{equation}
	\dot{x}(t)=-2y(t)Z,
	\label{xdot}
\end{equation}	
\begin{equation}
	\dot{y}(t)=2x(t)Z-2z(t)X
	\label{ydot}
\end{equation}	
and
\begin{equation}
	\dot{z}(t)=2y(t)X,
	\label{zdot}
\end{equation}	
where `$\cdot$' represents the derivative with respect to time. The mutual dependence of the above equations will help to express all the unknown arbitrary time-dependent functions $\left(x(t),y(t)\right.$ and $\left. z(t)\right)$ in terms of any one of these functions and its derivatives, which also gives a final condition that connects all of these functions. We formulate $x(t)$ and $y(t)$ from the equations (\ref{ydot}) and (\ref{zdot}) in terms of $z(t)$ and its derivatives as
\begin{equation}
	x(t)=\frac{\ddot{z}(t)}{4XZ}+\frac{z(t)X}{Z},
	\label{xt}
\end{equation}
\begin{equation}
	y(t)=\frac{\dot{z}(t)}{2X}.
	\label{yt}
\end{equation}
Rewriting the invariant $\mathcal{\hat{I}}$ in terms of the above functions, we obtain the explicit form, 
\begin{equation}
\mathcal{\hat{I}}=\left(\frac{\ddot{z}(t)}{4XZ}+\frac{z(t)X}{Z}\right)\sigma_{x}+\left(\frac{\dot{z}(t)}{2X}\right)\sigma_{y}+z(t)\sigma_{z},
\label{Invariantmain}
\end{equation}
 while equation~(\ref{xdot}) gives the necessary condition connecting all the control parameters of invariant $\mathcal{\hat{I}}$ and the LZ Hamiltonian. Expanding and integrating the equation (\ref{xdot}) using equations (\ref{xt}) and (\ref{yt}), we obtain the condition, 
 \begin{equation}
 	Z=\frac{\frac{\ddot{z}(t)}{4X}+z(t)X}{\left(A^{2}-\frac{\dot{z}^{2}}{4X^{2}}-z^{2}\right)^{\frac{1}{2}}},
 	\label{CondBzToZ}
 \end{equation}
 where `$A$' is a constant of integration. In the above formalism, $z(t)$ is the arbitrary function that can be engineered to achieve STA. This approach will use invariant for the unitary evolution in both EGC and EGE processes, keeping both the isochoric processes under respective $\mathcal{\hat{H}}_{LZ}$~(\ref{LZHamil}) as considered in the previous section. Thus, the set of eigenstates at the beginning and the end of the unitary processes should be shared by both $\mathcal{\hat{H}}_{LZ}$ and $\mathcal{\hat{I}}$, such that the cycle runs smoothly without a wavefunction collapse. The condition for sharing common eigenstates by invariant and LZ Hamiltonian is given by the commutation relation, $\left[\mathcal{\hat{I}},\mathcal{\hat{H}}_{LZ}\right]=0$. Substituting the particular form of the invariant $\mathcal{\hat{I}}$~(\ref{Invariantmain}) and LZ Hamiltonian in the above commutation relation, we end up in conditions,
\begin{equation}
	\dot{z}(0,\tau)=\ddot{z}(0,\tau)=0.
	\label{BCforz}
\end{equation}
Substitute these conditions to equation (\ref{CondBzToZ}) to obtain,
\begin{equation}
	z(0)=\frac{Az_{1}}{\sqrt{z_{1}^{2}+x^{2}}},
	\label{z0}
\end{equation}
\begin{equation}
	z(\tau)=\frac{Az_{2}}{\sqrt{z_{2}^{2}+x^{2}}},
	\label{ztau}
\end{equation}
where $z_{1},z_{2}$ and $x$ are particular values considered for LZ Hamiltonian in the previous section. We can set up any arbitrary functions $z(t)$ satisfying the above conditions for the invariant $\mathcal{I}$ to drive the qubit during EGC and EGE. The arbitrary selection of $z(t)$ determines the path of finite-time evolution, thus, affect the cost of implementation and other figures of performance of the engine. As the purpose this study is to analyze the dependence of performance on the cost of implementation, we arbitrarily choose,
\begin{equation}
	z(t)=z(0)+6\left(z(\tau)-z(0)\right) \left(\frac{t}{\tau}\right)^{5}+10\left(z(\tau)-z(0)\right)\left(\frac{t}{\tau}\right)^{3}-15\left(z(\tau)-z(0)\right)\left(\frac{t}{\tau}\right)^{4}.
	\label{zt}	
\end{equation}
From equations (\ref{z0}) to (\ref{zt}), it is clear that the protocol $z(t)$ changes for different values of $A$. The necessary relation (\ref{CondBzToZ}) for the invariance of $\mathcal{\hat{I}}$ can be constituted for different values of $A$. Thus, it is convenient to obtain different $Z$ functions from equation (\ref{CondBzToZ}) by substituting for the z(t) (\ref{zt}) and $X=x$ (\ref{Xeq}) for different values of $A$. This should be noted that it is possible to  obtain $X$ for $z(t)$ and any arbitrary $Z$. However, it might be a difficult task to find $X$ in terms of other parameters using equation (\ref{CondBzToZ}) (see Appendix B). Further, we can find $x(t)$ and $y(t)$ from equations (\ref{xt}) and (\ref{yt}) respectively to construct the particular functional form of $\mathcal{\hat{I}}$ in equation (\ref{Invariantmain}). In a nutshell, we are finding the parametric variation of $x(t), y(t)$ and $z(t)$ for different values of $A$ obeying the necessary condition in equation (\ref{CondBzToZ}). Then, the corresponding unitary evolution operator for the invariant dynamics is obtained by
\begin{equation}
U_{kl}=\mathcal{T}exp\left({-i\int_{0}^{\tau}\mathcal{\hat{I}}(t^{\prime}) dt^{\prime}}\right).
\label{UnitaryoperatorforI}
\end{equation} 
Further, the definitions and calculation method for work and efficiency of the engine are same as explained in section~\ref{NonAQOC}, but the fidelity will be replaced by the one given by the above evolution operator. This is how we consider the invariant based STA as a special case of non-adiabatic QHE. We use all the formalism developed in section \ref{NonAQOC} for a QHE with working medium described by the invariant $\mathcal{\hat{I}}$~(\ref{Invariantmain}). We can compare the use of resources for the dynamics with $\mathcal{\hat{H}}_{LZ}$ and $\mathcal{I}$ by calculating cost ratio, $\frac{\mathcal{C}_{I}}{\mathcal{C}_{LZ}}$. From equation (\ref{cost}), we write
	\begin{equation}
		\mathcal{C}_{I}\propto\int_{0}^{\tau} \vert\vert\mathcal{\hat{I}}\vert\vert^{2}dt
	\end{equation}
	and
	\begin{equation}
		\mathcal{C}_{LZ}\propto \int_{0}^{\tau}\vert\vert\mathcal{\hat{H}}_{LZ}\vert\vert^{2}dt.
	\end{equation} 
	By working out the above Frobenius norms for $\mathcal{I}$ and $\mathcal{\hat{H}}_{LZ}$, the ratio of the cost turns out to be
	\begin{equation}
		\frac{\mathcal{C}_{I}}{\mathcal{C}_{LZ}}=\frac{\int_{0}^{\tau}\left(x^{2}(t)+y^{2}(t)+z^{2}(t)\right)dt}{\int_{0}^{\tau}\left(Z^{2}+X^{2}\right)dt},
		\label{CReq}
	\end{equation}
	where the explicit $Z$ appearing in the above equation corresponding to $\mathcal{\hat{H}}_{LZ}$ given in the equation (\ref{Zeq}).
\begin{figure}[tb]
	\centering
	\begin{subfigure}{.49\linewidth}
		\includegraphics[scale=.8]{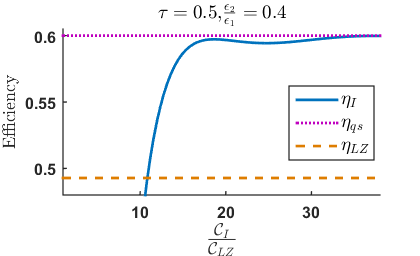}
		\subcaption{}
		\label{Effc54}
	\end{subfigure}
	\begin{subfigure}{.49\linewidth}
		\includegraphics[scale=.8]{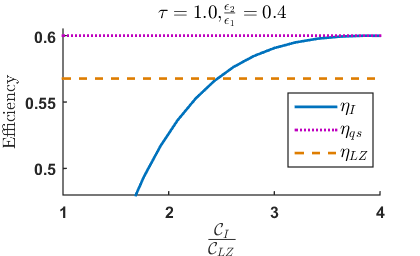}
		\subcaption{}
		\label{Effc14}
	\end{subfigure}
	\begin{subfigure}{.49\linewidth}
		\includegraphics[scale=.8]{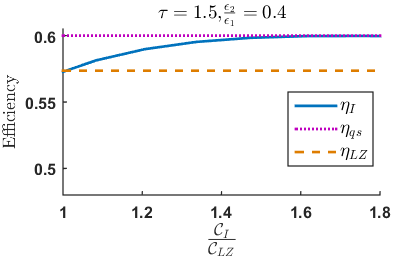}
		\subcaption{}
		\label{Effc154}
	\end{subfigure}
	\caption{The efficiency, $\eta_{I}$ of the invariant based QHE is plotted against the ratio of cost $\frac{\mathcal{C}_{I}}{\mathcal{C}_{LZ}}$ for the change in energy gap corresponding to the ratio $\frac{\epsilon_{2}}{\epsilon_{1}}=0.4$ with different time duration for the unitary processes, $\left(a\right)$ $\tau=0.5$, $\left(b\right)$ $\tau=1.0$ and $\left(c\right)$ $\tau=1.5$. Also, each plot indicate the corresponding efficiencies of LZ dynamics $\left(\eta_{LZ}\right)$ and quasi-static QHE $\left(\eta_{qs}\right)$ for comparison. $C_{I}$ is changed by the parametric variation of $x(t),y(t)$ and $z(t)$ with constant $X$ given by equations (\ref{xt}),(\ref{yt}),(\ref{zt}) and (\ref{Xeq}) respectively.}
	\label{EffcILZ1}
\end{figure}
\begin{figure}[tb]
	\centering
	\begin{subfigure}{.49\linewidth}
		\includegraphics[scale=.8]{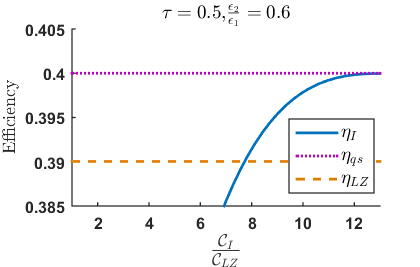}
		\subcaption{}
		\label{Effc56}
	\end{subfigure}
	\begin{subfigure}{.49\linewidth}
		\includegraphics[scale=.8]{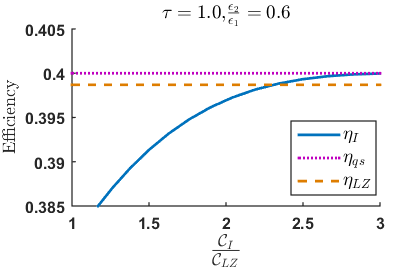}
		\subcaption{}
		\label{Effc16}
	\end{subfigure}
	\begin{subfigure}{.49\linewidth}
		\includegraphics[scale=.8]{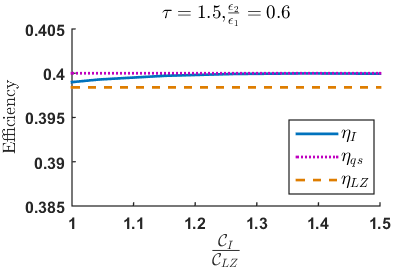}
		\subcaption{}
		\label{Effc156}
	\end{subfigure}
	\caption{The efficiency, $\eta_{I}$ of the invariant based QHE is plotted against the ratio of cost $\frac{\mathcal{C}_{I}}{\mathcal{C}_{LZ}}$, for the change in energy gap corresponding to the ratio $\frac{\epsilon_{2}}{\epsilon_{1}}=0.6$. The considered time duration of the unitary processes are $\left(a\right)$ $\tau=0.5$, $\left(b\right)$ $\tau=1.0$ and $\left(c\right)$ $\tau=1.5$. Also, each plot indicate the corresponding efficiencies of LZ dynamics $\left(\eta_{LZ}\right)$ and quasi-static QHE $\left(\eta_{qs}\right)$ for comparison. $C_{I}$ is changed by the parametric variation of $x(t),y(t)$ and $z(t)$ with constant $X$ given by equations (\ref{xt}),(\ref{yt}),(\ref{zt}) and (\ref{Xeq}) respectively.}
	\label{EffcILZ2}
\end{figure}

Figure~\ref{FI} shows the fidelity in the case of QHE with invariant based STA. From the above figure, we can analyze that the fidelity corresponding to LR invariant exceeds that of LZ Hamiltonian based QHE with few times extra cost than QHE with LZ Hamiltonian. Also, the fidelity gradually reaches unity with increasing cost ratio (see Appendix B for fidelity analysis of other two protocols for $z(t)$). We can expect a similar trend in the performance of the QHE engine with invariant based STA as the work is a linear function of fidelity~(see Eq. (\ref{workdone})). For comparison between the performance of STA enabled QHE and the LZ Hamiltonian based QHE, we have plotted the ratio of work $\frac{W_{I}}{W_{LZ}}$ for various time durations in Figure \ref{WR} and efficiency of the QHE with invariant dynamics in Figure \ref{EffcILZ1} and Figure \ref{EffcILZ2}. The subplots in each of the above figures show the variation of corresponding function (work ratio/efficiency) for a particular time duration with the changing cost ratio given in equation (\ref{CReq}). We have selected the lower time durations with more evident non-adiabaticity of LZ dynamics to study the work ratio and efficiency. Observing the work ratio plots, we can conclude that the required cost for improved performance of the engine ($W_{I}>W_{LZ}$) is much higher for a shorter time duration. Also, the performance difference is more perceptible at lower time duration with a larger change in the energy gap during the EGC and EGE ($\frac{\epsilon_{2}}{\epsilon_{1}}=0.4$). For the longer time duration, the work ratio approaches unity for both $\frac{\epsilon_{2}}{\epsilon_{1}}=0.4$ and $\frac{\epsilon_{2}}{\epsilon_{1}}=0.6$. This happens due to the gradual shift of LZ dynamics to its adiabatic nature with increasing time duration. Each of the efficiency plots at a particular time duration is combined with the corresponding efficiency of LZ dynamics. We can observe that invariant-based STA efficiency exceeds  LZ dynamics efficiency for all the selected lower time duration with a finite cost. As in the case of work ratio, the required cost is increasing with the lowering time duration of the process. If the change in the energy gap is small ($\frac{\epsilon_{2}}{\epsilon_{1}}=0.6$) during the EGC and EGE, the efficiency of invariant-based STA gets ahead in a much lower cost than that with bigger changes in the energy gap ($\frac{\epsilon_{2}}{\epsilon_{1}}=0.4$). We can also notice that the efficiency plots for $\tau=1.0$ shows the similar trend of the fidelity plot in Figure \ref{FI} due to the linear relation between the work and fidelity~(\ref{workdone}). Observing Figure \ref{WR}, Figure \ref{EffcILZ1} and Figure \ref{EffcILZ2}, It is worth mentioning that, for time duration $\tau=1.5$, the invariant dynamics shows triumph over the LZ dynamics with approximately equal cost ($\frac{\mathcal{C}_{I}}{\mathcal{C}_{LZ}}\approx 1$). Also, the quasi-static efficiency is achievable using our STA protocol in all the specified time durations with finitely higher cost than that of LZ Hamiltonian based QHE.

\section{Conclusion}
\label{Conclusion}   
The possibility of a non-adiabatic QHE with any arbitrary dynamics is predictable using the definitions of work and heat in connection with the fidelity of the unitary processes involved in the Otto cycle. The definition of work in this paper, as the sum of heat exchanged during the isochoric processes conventionally gives a positive value for a working QHE. Thus, the positive efficiency is a natural consequence of the above framework. The range of $\frac{\epsilon_{2}}{\epsilon_{1}}$ for a working QHE imply that the EGC process should follow thermalization with the hot bath to get the positive work output. The required fidelity for a QHE is not only constrained by the change in energy gap during the unitary processes but also by the temperature gradient between the hot and cold baths. We have analyzed the performance of QHE with LZ dynamics in terms of work and efficiency in the non-adiabatic regime. The characteristics of QHE with LZ Hamiltonian, $\mathcal{\hat{H}}_{LZ}$ tends to that of an adiabatic QHE at longer time durations. However, the amount of non-adiabaticity in shorter time durations results in decreased work and efficiency. We can reduce this non-adiabaticity and regain the lost efficiency of the engine by using the techniques of STA. We have followed the LR invariant-based STA method for QHE with LZ Hamiltonian and illustrated the performance's gradual improvement with the increasing cost of implementation. The cost is calculated by assuming electromagnetic fields for the construction of QHE, and using the method based on Frobenius norm of the Hamiltonian associated with the QHE. We have assigned the invariant dynamics only for EGC and EGE unitary processes, keeping the constant LZ Hamiltonian for the HI and CI. We found that the performance of the invariant-based QHE surpassed the QHE with LZ dynamics of the same cost but admitting the fact that the cost is several times greater for much lower time durations. This study of `cost' also implies that an invariant protocol is useful for STA process if and only if it is implemented with enough resources to overcome the non-adiabaticity in the reference system. The work and efficiency studies shows that the quasi-static performance of the QHE with LZ dynamics is achievable with the STA method illustrated in this paper. Apart from developing a STA protocol, the results of this paper suggest that it is equally important to find the cost, for which the protocol exhibit the required improvement in performance.
\section*{Appendix A: Time-dependent Schrodinger Equation; Numerical Approach}

	\begin{figure}[h]
		\centering
		\includegraphics[scale=0.3]{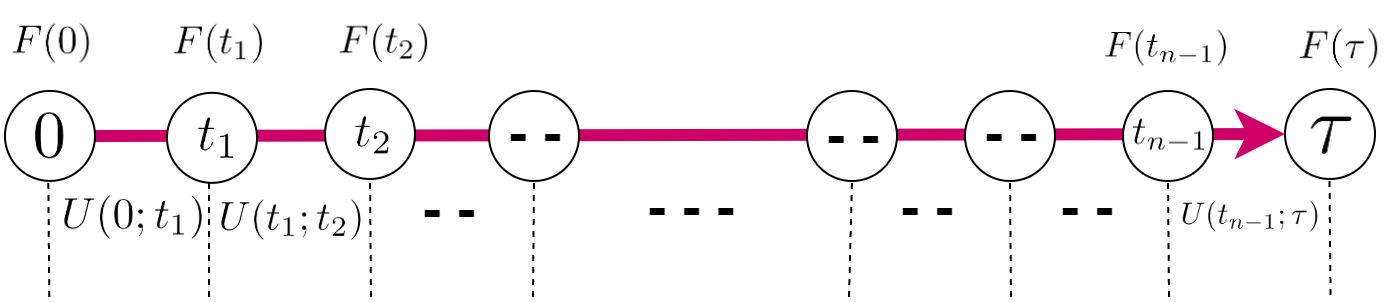}
		\caption{Splitting the total time duration in to $`n$' equal intervals gives $F(t)$ and the time evolution operators in a sequence. Algorithm to find the time evolution of $\tilde{\Psi}(t)$ should not change this sequence as the operator $F(t)$ at different instants of time do not commute in general $\left(\left[F(t_{a}),F(t_{b})\right]\neq0; a,b\in \left\{0,n\right\}\right)$.}
		\label{TimeEvolution}
	\end{figure} 
	Consider in general, this paper deals with the time evolution of some state $\tilde{\Psi}(t)$ corresponding to the Schrodinger equation $i\frac{\partial\tilde{\Psi}(t)}{\partial t}=F(t)\tilde{\Psi}(t)$. $F(t)$ is the generator of time evolution, which represents any of the Hermitian operators $\mathcal{\hat{H}}_{S}(t),\mathcal{\hat{H}}_{LZ}(t)$ and $\mathcal{\hat{I}}(t)$. Analytical solution might be existing for some particular $F(t)$. However, a numerical solution is the general possibility for any operator $F(t)$. Calculating the evolution of $\tilde{\Psi}(t)$ requires enormous computational power. The non-commuting $F(t)$ at different instants of time further restricts the algorithms to follow the $F(t)$ sequentially in time space. In other words, splitting the total time duration ($\tau$) in to `n' equal intervals gives a sequence of time evolution operators as explained in Figure \ref{TimeEvolution}. We can define $F(t_{i})$ at any instant of time $t_{i}$, where $i\in\left\{0,n\right\}$, the initial time $t_{0}=0$ and the final time $t_{n}=\tau$. By defining a step by step evolution operator $U(t_{i};t_{i+1})$, we can construct the total evolution operator as~\cite{SakuraiModern1994}
	\begin{equation}
		U(0;\tau)=\prod_{i=0}^{n-1}U(t_{i};t_{i+1}),
		\label{StepUniEvolution}
	\end{equation}
	where
	\begin{equation}
		U(t_{i};t_{i+1})=exp\left(-i\int_{t_{i}}^{t_{i+1}}F(t)dt\right).
	\end{equation}
	For a very large value of `n', we can approximate that the operator $F(t)=F\left(\frac{t_{i}+t_{i+1}}{2}\right)$ is fixed in between $t_{i}$ and $t_{i+1}$ for all $i\in\left\{0,n-1\right\}$~\cite{HansProduct1987,JinEigenstate2016}. The approximation of fixed generator for smaller intervals of time reduces the time evolution operator to
	\begin{equation}
		U(t_{i}:t_{i+1})=exp\left(-i\cdot F\left(\frac{t_{i}+t_{i+1}}{2}\right)\cdot(t_{i+1}-t_{i})\right).
	\end{equation}
	Iteratively finding $U\left(t_{i};t_{i+1}\right)$ gives the time evolution operator for the total duration, $U(0;\tau)$, which gives the final state $\tilde{\Psi}(\tau)$ from $\tilde{\Psi}(0)$ using the equation, $\tilde{\Psi}(\tau)=U(0;\tau)\tilde{\Psi}(0)$. The step by step algorithm is as follows,
	
	\textbf{Step 1:} Initialize the variables $n,t_{0}$ and $t_{n}$
	
	\textbf{Step 2:} Define the set of $n+1$ values in between $t_{0}$ and $t_{n}$
	
	\textbf{Step 3:} Iteratively find the values of $F\left(\frac{t_{i}+t_{i+1}}{2}\right)$ and $U(t_{i};t_{i+1})$ by looping over all the values of '$i$'.
	
	\textbf{Step 4:} Find $U(0;\tau)$, using the equation (\ref{StepUniEvolution}).
	
	\textbf{Stap 5:} Find $\tilde{\Psi}(\tau)$ from $\tilde{\Psi}(0)$ by applying $U(0;\tau)$.
	
	We have executed the above algorithm for $F(t)=\mathcal{\hat{H}}_{S}(t), \mathcal{\hat{H}}_{LZ}(t)$ and $\mathcal{\hat{I}}$. The initial time is set to 0 setting $\tau$ as the total duration of the process. Also, the algorithm is implemented for various $\tau$ values. We split the total time duration to 10000 small intervals ($n=10001$), which gave the precision up to three decimal points. This iterative approach always preserve the sequence of time evolution operator necessary for the non-commuting generators of time evolution. This method is partially inspired from the numerical approach explained in the Ref.~\cite{SchmidtkeStiffness2018}, which update the evolved state using the fourth-order Runge-Kutta iteration method. Instead of updating the evolved state, we update the evolution operator, which reduces the computational complexity by avoiding the use of fourth-order Runge-Kutta method for each loop of iteration.
\section*{Appendix B: Fidelity analysis for two more z(t) protocols}
\begin{figure}[h]
		\centering
		\begin{subfigure}{.49\linewidth}
			\includegraphics[scale=.6]{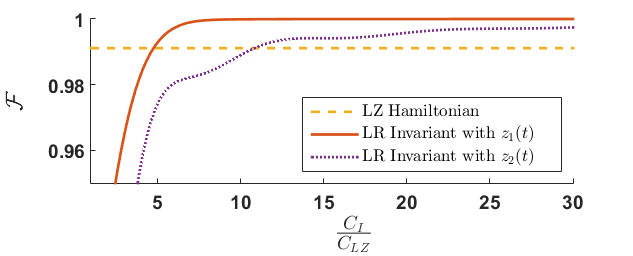}
			\subcaption{}
		\end{subfigure}
		\begin{subfigure}{.49\linewidth}
			\includegraphics[scale=.6]{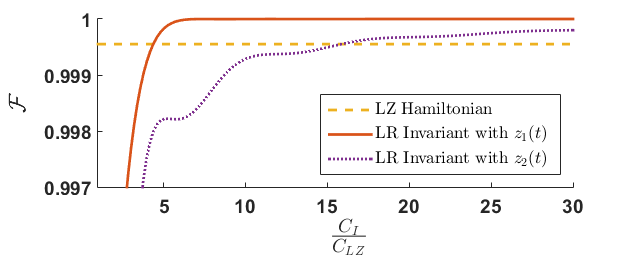}
			\subcaption{}
		\end{subfigure}
		\caption{Fidelity of the unitary processes using LR invariant is plotted against the cost ratio, $\frac{C_{I}}{C_{LZ}}$ for time duration, $\tau=1.0$. The corresponding fidelity of the QHE with LZ Hamiltonian is included in each plot for comparison. The change in the energy gap is fixed corresponding to the ratio, (a)$\frac{\epsilon_{2}}{\epsilon_{1}}=0.4$ and (b)$\frac{\epsilon_{2}}{\epsilon_{1}}=0.6$. $C_{I}$ is changed by the parametric variation of $x(t),y(t)$ and $z(t)$ (i.e., $z_{1}(t)$ or $z_{2}(t)$) with constant $X$ given by equations (\ref{xt}),(\ref{yt}),(\ref{z1t}),(\ref{z2t}) and (\ref{Xeq}) respectively.}
	\label{FApndx}
\end{figure}
	We have been using the protocol given in equation (\ref{zt}) to analyze the cost and performance of invariant based STA enabled QHE. A natural question arises is the variation in efficiency and work for some other protocol for $z(t)$. The dependence of performance on the selection of $z(t)$ is completely arbitrary resulting from the random path of evolution. However, we can analyze such variations in performance by constituting different protocols for $z(t)$ obeying the conditions given in equations (\ref{BCforz}),(\ref{z0}) and (\ref{ztau}). In this appendix, we consider two more feasible protocols for $z(t)$,
\begin{equation}
	z_{1}(t)=z(0)+\left(z(\tau)-z(0)\right)sin^{2}\left[\frac{\pi}{2}sin^{2}\left(\frac{\pi t}{2\tau}\right)\right],
	\label{z1t}
\end{equation} 
\begin{equation}
	z_{2}(t)=z(0)+30\left(z(\tau)-z(0)\right) \left(\frac{t}{\tau}\right)^{4}-54\left(z(\tau)-z(0)\right)\left(\frac{t}{\tau}\right)^{5}+25\left(z(\tau)-z(0)\right)\left(\frac{t}{\tau}\right)^{6},
	 \label{z2t}
\end{equation} 
where, $z_{1}(t)$ is inspired from Ref.~\cite{AndreasMany2020} and we have constructed $z_{2}(t)$ by our own. Although it is labeled as $z_{1}(t)$ and $z_{2}(t)$ for distinguishability, but it follows all the necessary conditions for $z(t)$ and can replace $z(t)$ in section (\ref{STAI}) to analyze the performance of the engine. 
	
	The performance analysis in section \ref{STAI}, shows the linear dependence of work and efficiency on the fidelity of unitary process (i.e.,  the  heat exchanged, that characterizing the heat engine is a linear function of fidelity). Thus, the comparison of fidelity among LZ dynamics and Invariant dynamics of the engine gives intuitive conclusion regarding the performance of the engine. Figure \ref{FApndx} shows the fidelity of invariant-based STA engine with protocol $z_{1}(t)$ and $z_{2}(t)$ along with fidelity of LZ Hamiltonian for comparison. By analyzing the fidelity plots and comparing with figure \ref{FI}, we can conclude that, the both the protocols $z_{1}(t)$ and $z_{2}(t)$ cost more than that of the protocol in equation (\ref{zt}) to outperform the LZ dynamics. Also, the arbitrariness in dependence of fidelity (equivalently performance) on different protocol is evident from the plot.

\end{document}